\documentclass[conference,a4paper,table]{APSIPA2025}
\usepackage{amsmath}
\usepackage{graphicx}
\usepackage{multirow}
\usepackage{multicol}
\usepackage{threeparttable}
\usepackage{booktabs}

\usepackage{geometry}
\usepackage{fancyhdr}

\fancypagestyle{firststyle}{
  \fancyhf{}
  \fancyhead[C]{2025 Asia Pacific Signal and Information Processing Association Annual Summit and Conference (APSIPA ASC)}
}

\begin{document}

\title{Fusion of Modulation Spectrogram and SSL with Multi-head Attention for Fake Speech Detection}
\author{
\begin{tabular}{c}
Rishith Sadashiv T N\textsuperscript{1} \quad
Abhishek Bedge\textsuperscript{1} \quad
Saisha Suresh Bore\textsuperscript{2} \quad
Jagabandhu Mishra\textsuperscript{3} \\
Mrinmoy Bhattacharjee\textsuperscript{4} \quad
S R Mahadeva Prasanna\textsuperscript{1,2} \\[0.2em]
\textsuperscript{1}IIT Dharwad \quad
\textsuperscript{2}IIIT Dharwad \quad 
\textsuperscript{3}University of Eastern Finland\quad
\textsuperscript{4}IIT Jammu \\[0.2em]
Email: ee24dpp10@iitdh.ac.in
\end{tabular}
}

\maketitle
\thispagestyle{firststyle}
\pagestyle{fancy}

\begin{abstract}
  Fake speech detection systems have become a necessity to combat against speech deepfakes. Current systems exhibit poor generalizability on out-of-domain speech samples due to lack to diverse training data. In this paper, we attempt to address domain generalization issue by proposing a novel speech representation using self-supervised (SSL) speech embeddings and the Modulation Spectrogram (MS) feature. A fusion strategy is used to combine both speech representations to introduce a new front-end for the classification task. The proposed SSL+MS fusion representation is passed to the AASIST back-end network. Experiments are conducted on monolingual and multilingual fake speech datasets to evaluate the efficacy of the proposed model architecture in cross-dataset and multilingual cases. The proposed model achieves a relative performance improvement of 37\% and 20\% on the ASVspoof 2019 and MLAAD datasets, respectively, in in-domain settings compared to the baseline. In the out-of-domain scenario, the model trained on ASVspoof 2019 shows a 36\% relative improvement when evaluated on the MLAAD dataset. Across all evaluated languages, the proposed model consistently outperforms the baseline, indicating enhanced domain generalization. 
\end{abstract}

\section{Introduction} \label{sec:introduction}

In recent years, the sophistication of machine-generated speech has increased significantly, enabling both beneficial and malicious applications. While generative speech technology supports valuable use cases such as assistive tools and accessibility, it also poses serious threats when misused—for instance, in spreading manipulated war narratives or deceiving speaker verification (SV) systems. The rapid progress in this field introduces ongoing challenges for designing effective countermeasure systems, particularly for \emph{Fake Speech Detection (FSD)}. FSD has been extensively studied, evolving from the use of hand-crafted features and simple classifiers to end-to-end deep neural networks like RawNet2~\cite{fsd_rawnet2} and AASIST~\cite{aasist_original}. More recently, SSL models and state-space architectures like Mamba have shown promise~\cite{baseline_sslAasist,xiao2025xlsr}. For real-time deployment, systems trained in one domain must generalize well to others. However, domain generalizability remains a persistent challenge, as FSD models often struggle to maintain performance across datasets due to variations in recording conditions and dataset-specific characteristics.

\begin{figure}[t]
    \begin{center}
        \includegraphics[width=0.86\linewidth]{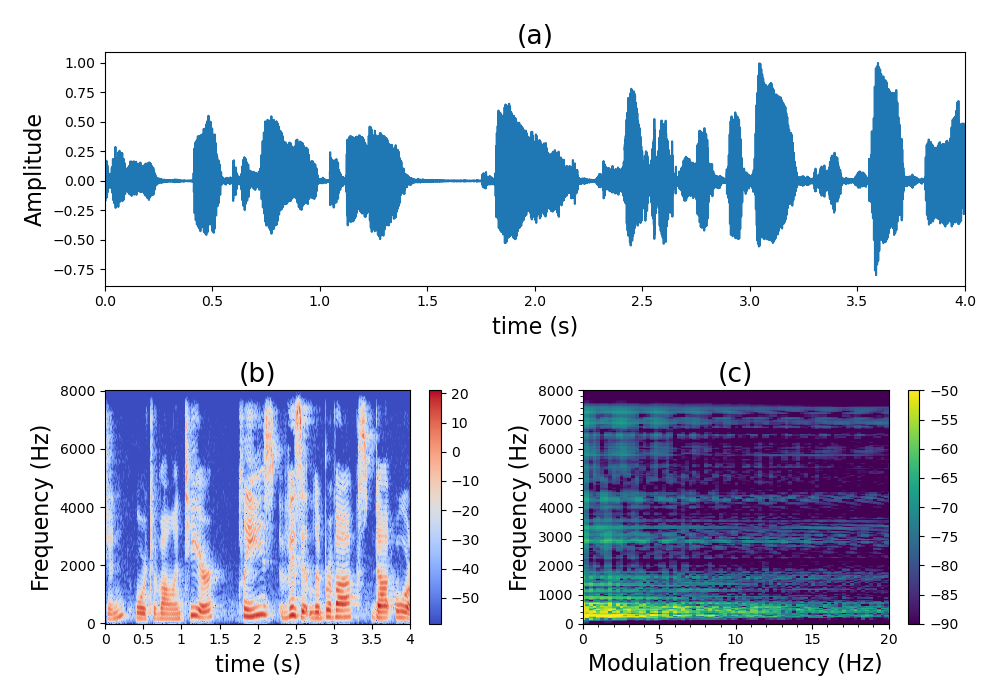} 
    \end{center}
    \vspace{-1.4em}
    \caption{(a) Speech signal, (b) Spectrogram, and (c) Modulation spectrogram}
    \label{fig:ms}
    \vspace{-1.8em}
\end{figure}

\begin{figure*}[t]
    \begin{center}
        \includegraphics[width=0.8\linewidth]{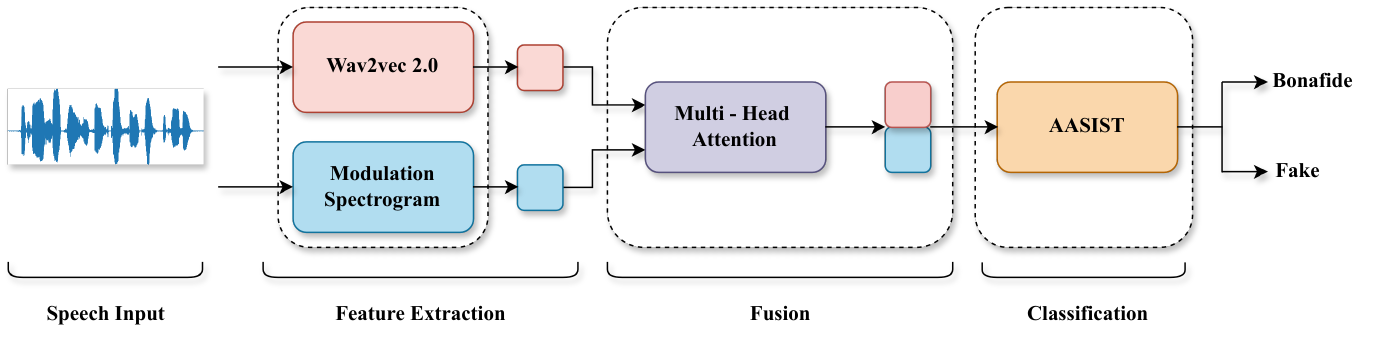}
    \end{center}
    \vspace{-1.5em}
    \caption{ \centering Proposed methodology: In the fusion block, wav2vec 2.0 embeddings form the key and value, with the modulation spectrogram as the query. }
    \label{fig:blockDiagram}
    \vspace{-1.0em}
\end{figure*}

Many works have established the performance degradation of FSD systems in out-of-domain scenarios~\cite{crossDb_empiricalStudy,das2020assessing}. To resolve this, the attempts are broadly in two directions to improve generalization ability of FSD system, (1) use of specialized model training strategies, and (2) signal processing and data-driven approaches. Various training strategies have been explored to improve generalization, including multi-task meta-learning~\cite{meta-learning}, continual learning~\cite{continualLearning}, one-class learning~\cite{oneClassLearning}, and optimal transport-based domain adaptation~\cite{optimalTransport}. Many existing approaches focus on signal processing techniques to extract novel features for FSD. For instance, the study in~\cite{crossDb1} proposed the application of 2D Discrete Cosine Transform (2D-DCT) on log-Mel spectrograms. Pronunciation and prosodic features have also been explored in~\cite{fusion_referencePaper} to enhance generalization. Furthermore, a combination of modulation spectrogram and residual modulation spectrogram features has been investigated in~\cite{fsd_ms_rms}. SSL front-ends have gained popularity in recent years. The study in~\cite{wav2vec2.0_finetuning} demonstrated that fine-tuning the wav2vec 2.0 XLS-R model on an FSD dataset leads to improved domain generalization, even when paired with a simple fully connected (FC) back-end. Furthermore, the results indicate that the wav2vec 2.0 model provides better FSD generalization compared to other SSL models like HuBERT. Similarly, another work~\cite{crossDb3} investigates the use of a variational information bottleneck module along with a wav2vec-based front-end and an FC back-end. However, we hypothesize that a representation derived by combining signal processing-based features with data-driven SSL embeddings could potentially be a promising approach for the FSD task.

In this study, we propose a \emph{novel front-end representation for improved domain generalization} in the FSD task. We achieve this by \emph{combining} wav2vec 2.0 cross-lingual self-supervised speech representations (\emph{XLS-R}), which is hereafter referred to as the SSL model, with the \emph{modulation spectrogram} feature. While the modulation spectrogram has been previously introduced for FSD in~\cite{fsd_ms_rms}, and SSL model embeddings have been combined with other speech features to enhance generalizability in~\cite{fusion_referencePaper}, to the best of our knowledge, the joint use of modulation spectrogram and SSL embeddings has not yet been explored. To address this gap, we employ a multi-head attention mechanism as the fusion strategy. Since SSL models are primarily trained to capture speech characteristics at the word or syllable level, they may not effectively represent frame-level artifacts. In contrast, modulation spectrogram provides variation in speech dynamics from frame-level to prosodic level. Fig.~\ref{fig:ms} illustrates the modulation spectrogram feature alongside the corresponding speech waveform and spectrogram. We hypothesize that the fusion of SSL embeddings with modulation spectrogram features can yield a more generalizable representation for FSD. The AASIST network is employed as the back-end architecture, as described in~\cite{baseline_sslAasist}. The effectiveness of the proposed system is evaluated on the monolingual ASVspoof 2019 Logical Access (LA) dataset, followed by domain generalization experiments using the recent multilingual MLAAD dataset~\cite{mlaad_orig}. Additionally, the impact of language variation is analyzed using the MLAAD dataset. Experimental results demonstrate that the proposed fusion-based front-end significantly enhances domain generalization compared to the baseline. The main contributions of this paper are summarized as follows:

\begin{itemize}
    \item We propose the fusion of modulation spectrogram feature with SSL model embeddings for the FSD task.
    \item A novel architecture is introduced that employs the fused feature representation as the front-end and the AASIST network as the back-end.
    \item Validation of the proposed framework on cross-domain and multi-lingual setup.
\end{itemize}

The remainder of the paper is organized as follows:  Section~\ref{sec:methodology} illustrates the proposed methodology. The experimental setup is described in Section~\ref{sec:experimental_setup}. The results are reported in Section~\ref{sec:results} along with discussions. The paper is concluded in Section~\ref{sec:conclusion}.

\section{Methodology} \label{sec:methodology}
In this section, we describe our proposed approach for the FSD task, which fuses SSL features with modulation spectrogram using multi-head attention. Fig.~\ref{fig:blockDiagram} illustrates the overall architecture. Building on the widespread use of Audio Anti-Spoofing using Intgrated Spectro-Temporal graph attention networks (AASIST) with SSL features in prior work~\cite{aasist_original, baseline_sslAasist}, we combine the fused representation with AASIST to perform spoofing detection. In the following subsections, we briefly explain the modulation spectrogram, SSL features, the fusion process using multi-head attention, and the AASIST model.

\subsection{Modulation Spectrogram.} The modulation spectrogram provides a two-dimensional representation of a speech signal. To compute it, we follow a two-step process. First, we apply a Short-Term Fourier Transform (STFT) to the speech signal $x(t)$ to obtain the spectrogram $X(t,f)$, which serves as a time-frequency representation. The frequency $f \in [0, f_N]$ and time $t \in [0, T]$, where $N$ and $T$  denote the number of FFT points and the total number of time samples, respectively. Next, we compute the modulation spectrogram $Y(f_{mod}, f)$ by applying a Fourier transform over time to the magnitude of each frequency component of $X(t,f)$. This transformation yields:
\begin{equation} \label{eq:ms}
    Y(f_{mod}, f_i) = \mathcal{F}{\{|X(t,f_i)|\}}, i=0 \ldots f_N
\end{equation}
where $\mathcal{F}$ denotes the Fourier transform. We use $\hat{N}$ FFT points for modulation spectrogram computation, which is equal to the number of frames in the STFT. The resulting modulation spectrogram captures the conventional frequency 
$f$ along one axis and the modulation frequency 
$f_{mod}$ along the other~\cite{ms_amaPaper}. The modulation frequency $f_{mod}$  captures how the temporal dynamics of the speech signal vary, from rapid changes at the frame level to slower trends at the prosodic level.

\subsection{SSL Embeddings}

We use the XLS-R variant of the wav2vec 2.0 model~\cite{wav2vec2.0_orig, wav2vec2.0_xls-r} to extract feature representations from speech signals. The model employs a multi-layer convolutional neural network as a feature encoder $f:\mathcal{X} \rightarrow \mathcal{Z}$, which transforms raw waveforms $x_{1:T}$ into latent speech representations $z_{1:\hat{T}}$ where $T$ denotes the number of samples and $\hat{T}$ represents the number of time steps. The stride of the feature encoder determines the value of $\hat{T}$. The model then feeds the speech representations into transformer network $g: \mathcal{Z} \rightarrow \mathcal{C}$, which produces context representations $c_{1:\hat{T}}$ that capture information from the entire latent representation sequence in an end-to-end manner.

During self-supervised training, the latent speech representations $z_{1:\hat{T}}$ are quantized to a finite set of speech representations $q_{1:\hat{T}}$ using a quantization module $\mathcal{Z} \rightarrow \mathcal{Q}$. It involves quantized representations from multiple codebooks. The latent representations are masked at random starting points before being fed to the transformer network. The model training involves solving a contrastive task, which requires identifying the true quantized latent representation $q_t$ for a masked time step within a set of distractors. The contrastive loss is augmented with a codebook diversity loss to encourage the model to use all codebook entries. The XLS-R model with $0.3$ billion parameters available at Fairseq toolkit~\cite{fairseq} is used in our study.

\subsection{Fusion Strategy using Multi-Head Attention}

We perform the fusion of XLS-R embeddings and modulation spectrogram using a multi-head attention network~\cite{fusion_referencePaper}. The multi-head attention mechanism conducts multiple scaled dot-product attention operations, as defined in~(\ref{eq:attention}):
\begin{equation} \label{eq:attention}
Attention (Q, K, V) = Softmax\left(\frac{KQ^T}{\sqrt{d_k}}\right)V,
\end{equation}
where we use the XLS-R embeddings as the key ($K$) and value ($V$), and the modulation spectrogram feature as the query ($Q$). The key and query both have dimensionality $d_k$, and the value has dimensionality $d_v$. We perform projection operations using multiple FC layers to generate $h$ sets of ($Q$, $K$, $V$) representations. We apply the attention operation to each set in parallel. Then, we concatenate the resulting outputs from all attention heads and project them through an FC layer~\cite{attention_paper}, as shown in, 
\begin{equation} \label{eq:multi-head_attention}
    \begin{split}
        MultiHead(Q, K, V) = Concat(head_1,\cdots, head_h)W^O \\ head_i = Attention(QW_i^Q, KW_i^K, VW_i^V)
    \end{split}
\end{equation}
where $W_i^Q$, $W_i^K$, $W_i^V$, and $W_i^O$ denote the parameter matrices of the FC layers for the $i^\text{th}$ head, and $i \in [1, h]$.

\subsection{AASIST Spoofing Detection}
AASIST is a widely used graph neural network framework for FSD. It uses a sinc-convolution layer to extract front-end features from raw audio, which are then encoded by a RawNet2 variant~\cite{fsd_rawnet2}. The model reshapes the output into a 2D representation and passes it through six residual blocks to extract high-level features. Two parallel graph modules, each with graph attention and pooling layers, model spectral and temporal artifacts. A max graph operation combines their outputs using two heterogeneous graph branches, followed by an element-wise maximum. Each branch uses two HS-GAL layers and pooling, with a stack node aggregating information. Final output uses max and average pooling, followed by a two-node output layer~\cite{aasist_original}. In~\cite{baseline_sslAasist}, wav2vec 2.0 replaces the sinc-convolution layer, and its embeddings are input to RawNet2. In our method, we replace wav2vec 2.0 with fused features, which are passed to the AASIST back-end.

\section{Experimental Setup} \label{sec:experimental_setup}

\begin{table}[t]
\begin{center}
\begin{threeparttable}
\caption{Composition of the datasets}
\begin{tabular}{ll|lll}
\toprule
Dataset &  & \begin{tabular}[l]{@{}c@{}}ASVspoof \\ 2019 LA\end{tabular} & \begin{tabular}[l]{@{}c@{}}ASVspoof\\ 2021 LA\end{tabular} & MLAAD \\ \midrule
\multirow{2}{*}{Train} & Bonafide & 2580 & - & 28345 \\  
  & Fake & 22800 & - & 36566 \\ \midrule
\multirow{2}{*}{Development} & Bonafide & 2548 & - & 6584  \\  
 & Fake & 22296 & - & 9765  \\ \midrule
\multirow{2}{*}{Evaluation}  & Bonafide & 7355 & 14816 & 6390  \\  
 & Fake & 63882 & 133360 & 19675 \\ \bottomrule
\end{tabular} \label{tab:dataset_composition}
\end{threeparttable}

\end{center}
\vspace{-2.6em}
\end{table}

\subsection{Datasets}
We use three datasets in this work: ASVspoof 2019, ASVspoof 2021, and MLAAD. We choose ASVspoof 2019 due to its widespread use in the literature, ASVspoof 2021 to assess generalizability under channel and noise variations, and MLAAD to evaluate generalization across different languages. Table~\ref{tab:dataset_composition} summarizes the key characteristics of these datasets, and the following subsections provide a brief description of each.

\subsubsection{ASVspoof 2019}
We use the LA partition of the ASVspoof 2019 dataset, which includes fake speech samples generated using $17$ different neural acoustic and waveform-based TTS and VC spoofing techniques. The dataset is divided into three subsets: train, development, and evaluation, each containing non-overlapping speakers. The train and development sets include fake speech samples from $6$ known spoofing techniques, while the evaluation set contains samples from $13$ spoofing techniques, $2$ known and $11$ unknown. The bonafide (genuine) speech samples come from the VCTK corpus~\cite{asvspoof2019}. This dataset is monolingual and includes only English-language speech.

\subsubsection{ASVspoof 2021} 
The ASVspoof 2021 dataset provides an updated evaluation set with an increased number of bonafide and fake speech samples. Unlike ASVspoof 2019, which contains studio-quality recordings, the 2021 evaluation samples are passed through telephony systems (VoIP and PSTN) to simulate real-world, in-the-wild conditions~\cite{asvspoof2021_2023}. This dataset is also monolingual and contains only English-language speech.

\subsubsection{MLAAD} 
The MLAAD dataset contains fake speech samples generated in $23$ different languages using $52$ state-of-the-art models across $22$ architectures~\cite{mlaad_orig}. It builds on the M-AILABS speech dataset~\cite{m-ailabs}, which provides bonafide speech in $8$ European languages. For languages not covered by M-AILABS, English text is translated into the target languages and then synthesized into fake speech using various TTS models sourced from \emph{Coqui.ai}\footnote{https://github.com/coqui-ai/TTS} and \emph{Hugging Face}. Following the protocols from~\cite{mlaad_protocol}, we split the dataset into training, development, and evaluation subsets with no speaker overlap. The bonafide samples come from $5$, $4$, and $4$ languages in the train, development, and evaluation sets respectively, while the fake samples include all $23$ languages across each subset.

\subsection{Evaluation Metric}
We use Equal Error Rate (EER) as the metric throughout this work. It represents the threshold at which the false alarm rate and miss rate are approximately equal, as shown in~(\ref{eq:eer}). 

\begin{equation}
    \label{eq:eer}
    EER = P_{fa}^{cm}(\tau_{EER}) \approx P_{miss}^{cm}(\tau_{EER})
\end{equation}

\subsection{Modulation Spectrogram Extraction}
We restrict all audio samples in this work to approximately $4$ seconds with a sampling rate of $16$ kHz ($64,600$ samples). We zero-pad shorter audios to match this length. Then, we extract the modulation spectrogram feature from the speech signal using a frame length of $25$ ms and a frame shift of $10$ ms. For STFT computation, the number of FFT points is set equal to the window length, i.e., $0.025 \times 16000=400$. For modulation spectrogram computation, the number of FFT points ($\hat{N}$) is set to the number of STFT frames, i.e., 402. This results in a modulation spectrogram feature dimension of $(\frac{N}{2}+1)\times(\frac{\hat{N}}{2}+1)=201\times202$.

\subsection{Fusion using multi-head attention}
The fusion operation follows a similar approach to that described in~\cite{fusion_referencePaper}.
The self-supervised XLS-R ($0.3$B) model\footnote{https://github.com/facebookresearch/fairseq/tree/main/examples/wav2vec} is used as SSL model. Raw audio segments of approximately 4 seconds are input to the SSL model, producing embeddings of size $201\times1024$. These embeddings are subsequently projected to a lower dimension of $201\times 128$ via an FC layer. For the fusion strategy, the key and value representations are obtained by passing the SSL embeddings through two separate FC layers, while the query is derived from the modulation spectrogram feature using another FC layer. These components are then processed by a multi-head attention block with $h=4$ heads. The output of the attention block is further projected through a final FC layer. All FC layers used in the fusion module output a fixed dimension of $P=256$, resulting in a final fused representation of size $201\times P$. This fusion representation is then fed into the AASIST back-end network. The entire model, including the SSL front-end, fusion module, and back-end, is jointly optimized during training.

\subsection{Training Details}
We apply RawBoost data augmentation on the fly to the existing training data using the same parameters and configuration as the baseline work~\cite{baseline_sslAasist}. 
We use a batch size of $14$ and a fixed learning rate of $10^{-6}$. 
We optimize the model with the standard Adam optimizer and use a weighted cross-entropy loss.
All models are trained for 100 epochs on an A100 GPU and we choose the model with the best development loss for testing.
The implementation is available in the Github repo\footnote{https://github.com/rishithSadashiv/ssl-ms-fsd}.

\begin{table}[t]
\caption{Cross-dataset EER (\%) comparison between the baseline (SSL) and the proposed fusion (SSL+MS) architectures. The LA part of the ASVspoof datasets has been used. Combined denotes ASVspoof 2019 + MLAAD datasets.}
\begin{center}
\begin{threeparttable}
\scalebox{0.88}{
\begin{tabular}{l|lllll}
\toprule
 \multirow{2}{*}{Model} &  & \multicolumn{3}{l}{Test Set $\rightarrow$} \tabularnewline \cmidrule{3-5} 
  & Train Set $\downarrow$ &  ASVspoof 2019  & ASVspoof 2021 & MLAAD  \\ \midrule
 \multirow{3}{*}{SSL}  & \cellcolor[HTML]{c1dbf7} ASVspoof 2019   & \cellcolor[HTML]{c1dbf7} 0.27 & \cellcolor[HTML]{c1dbf7} 1.02 & \cellcolor[HTML]{c1dbf7} 27.97  \\ 
    & \cellcolor[HTML]{e2d7fa} MLAAD & \cellcolor[HTML]{e2d7fa} 38.49 & \cellcolor[HTML]{e2d7fa} 37.85 & \cellcolor[HTML]{e2d7fa} 8.24 \\         
    & \cellcolor[HTML]{fac0be} Combined & \cellcolor[HTML]{fac0be} 1.33 & \cellcolor[HTML]{fac0be} 15.09 & \cellcolor[HTML]{fac0be} 9.72 \\ \midrule
 \multirow{3}{*}{SSL+MS} & \cellcolor[HTML]{c1dbf7}  ASVspoof 2019 & \cellcolor[HTML]{c1dbf7} 0.17 & \cellcolor[HTML]{c1dbf7} 1.15 & \cellcolor[HTML]{c1dbf7} 17.89 \\         
    & \cellcolor[HTML]{e2d7fa} MLAAD & \cellcolor[HTML]{e2d7fa} 40.89 & \cellcolor[HTML]{e2d7fa} 48.45 & \cellcolor[HTML]{e2d7fa} 6.52 \\
    & \cellcolor[HTML]{fac0be} Combined & \cellcolor[HTML]{fac0be} 0.34 & \cellcolor[HTML]{fac0be} 3.04 & \cellcolor[HTML]{fac0be} 5.79 \\
\bottomrule
\end{tabular}}
\end{threeparttable}    
\end{center}
\label{tab:results_table}
\vspace{-2.6em}
\end{table}

\begin{figure*}[t]
    \begin{center}
        \includegraphics[width=0.915\linewidth]{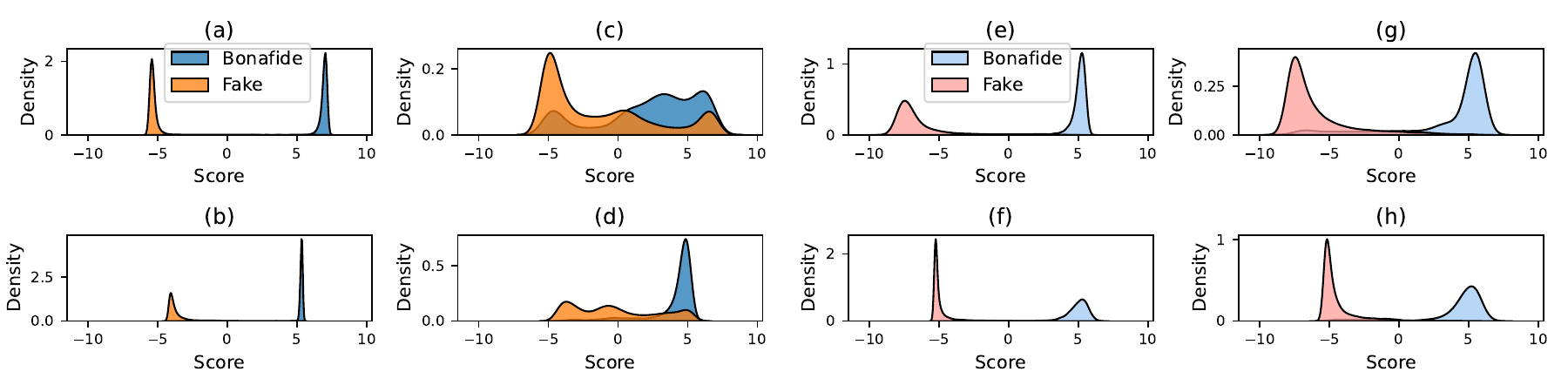}
    \end{center}
    \vspace{-1.6em}
    \caption{Score density plots for bonafide and fake speech across different datasets. Plots (a)–(d) show models trained on the ASVspoof 2019 dataset, while plots (e)–(h) show models trained on the combined ASVspoof 2019 and MLAAD datasets. (a) Baseline model on ASVspoof 2019 eval set (0.27\% EER), (b) Fusion model on ASVspoof 2019 eval set (0.17\% EER), (c) Baseline model on MLAAD eval set (27.97\% EER), (d) Fusion model on MLAAD eval set (17.89\% EER), (e) Baseline model on ASVspoof 2019 eval set (1.33\% EER), (f) Fusion model on ASVspoof 2019 eval set (0.34\% EER), (g) Baseline model on MLAAD eval set (9.72\% EER), (h) Fusion model on MLAAD eval set (5.79\% EER).}
    \label{fig:kde_plots}
    \vspace{-1.0em}
\end{figure*}

\begin{figure}[t]
    \begin{center}
        \includegraphics[width=0.88\linewidth]{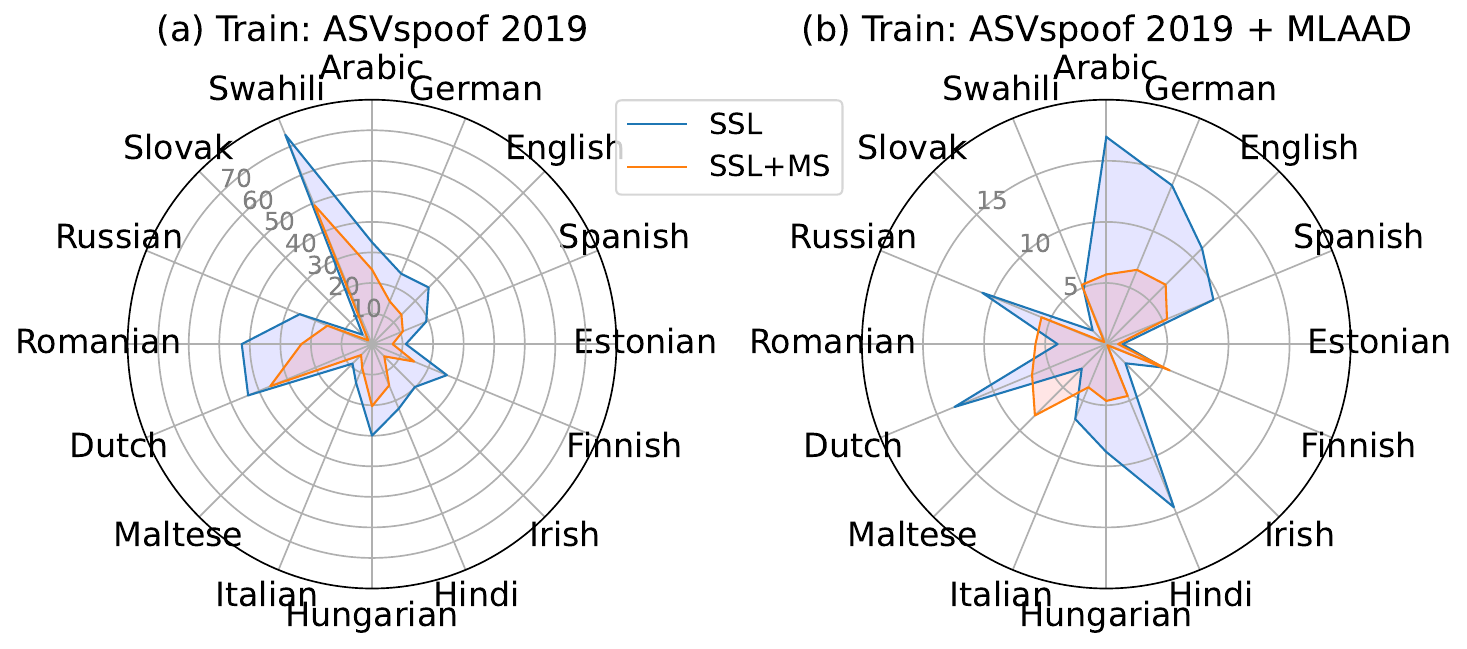} 
    \end{center}
    \vspace{-1.0em}
    \caption{Radar plots of EER (\%) for baseline (SSL) and proposed (SSL+MS) models across languages in the MLAAD evaluation set. (a) Out-of-domain case: models are trained using monolingual ASVspoof 2019 LA dataset (Overall performance: SSL: $27.97$\% EER and SSL+MS: $17.89$\% EER). (b) Models are trained on combination of ASVspoof 2019 LA + MLAAD dataset (Overall performance: SSL: $9.72$\% EER and $5.79$\% EER).}
    \label{fig:radar_plot}
    \vspace{-1.9em}
\end{figure}

\section{Results and Discussions} \label{sec:results}
This section reports the results of the baseline and proposed fusion models. We use SSL-AASIST~\cite{baseline_sslAasist} as the baseline and denote the proposed fusion model of SSL and modulation spectrogram as (SSL+MS)-AASIST. We train models on ASVspoof 2019, MLAAD, and their combination, and evaluate them on the ASVspoof 2019, ASVspoof 2021, and MLAAD evaluation sets. Table~\ref{tab:results_table} presents the performance across all evaluation scenarios. 

\subsection{Baseline: SSL with AASIST}
The baseline model trained on the ASVspoof 2019 dataset achieves strong in-domain performance with an EER of $0.27\%$ and generalizes reasonably well to the ASVspoof 2021 dataset, where it reaches an EER of $1.02\%$. However, it performs poorly on the out-of-domain MLAAD dataset, yielding a high EER of $27.97\%$. When trained on the MLAAD dataset, the model records an in-domain EER of $8.24\%$ but fails to generalize, with EERs of $38.49\%$ on ASVspoof 2019 and $37.85\%$ on ASVspoof 2021. In comparision to the in-domain performance, training the model on the combined ASVspoof 2019 and MLAAD datasets leads to a slight degradation in ASVspoof 2019 performance (EER of $1.33\%$), a substantial drop in ASVspoof 2021 performance (EER of $15.09\%$), and a moderate decrease on MLAAD (EER of $9.72\%$). These results show that although the baseline model performs well in in-domain settings, it struggles to generalize across domains, highlighting the impact of dataset-specific characteristics on model performance.

\subsection{Proposed: Fusion of Modulation spectrogram and SSL embeddings with AASIST} 

We conducted the same set of cross-domain experiments using the proposed fusion model. Notably, the fusion model achieves improved in-domain results, with an EER of $0.17\%$ on ASVspoof 2019 and $6.52\%$ on MLAAD, outperforming the corresponding baseline models. In out-of-domain evaluations, the ASVspoof 2019-trained fusion model performs comparably to its baseline counterpart on the ASVspoof 2021 dataset and shows enhanced performance on MLAAD. In contrast, the MLAAD-trained fusion model continues to perform poorly on both ASVspoof datasets, mirroring the trend observed in the baseline. This may be attributed to insufficient number of English language speech samples in MLAAD dataset.

Interestingly, the fusion model trained on the combined ASVspoof 2019 and MLAAD datasets demonstrates significant improvements. While its performance on ASVspoof 2019 and ASVspoof 2021 slightly lags behind the ASVspoof 2019-only trained fusion model, it clearly outperforms the baseline across all evaluation sets. On the MLAAD dataset, it even surpasses the in-domain performance of the MLAAD-trained fusion model. These results suggest that incorporating diverse data during training enables the fusion model to learn broader feature representations, which improves its generalizability and robustness to domain shifts.

Fig.~\ref{fig:kde_plots} presents density plots of classification scores from different models across various evaluation sets. The top row corresponds to the baseline SSL model, and the bottom row represents the proposed fusion (SSL+MS) model. Plots (a) and (b), which show in-domain results on ASVspoof 2019, reveal clean score separation for both models. However, in the out-of-domain case (plots c and d), where models are trained on ASVspoof 2019 and evaluated on MLAAD, the fusion model shows narrower bonafide score distribution, indicating improved domain generalization despite high EER values ($27.97\%$ for baseline vs. $17.89\%$ for fusion).

In the last four plots (e–h), we repeat the experiment using the combined training set. Comparing these plots clearly shows that the fusion model consistently achieves better separation between bonafide and fake scores than the baseline, reflecting the EER trends reported in Table~\ref{tab:results_table}. In summary, the proposed fusion architecture not only enhances in-domain performance but also significantly improves generalization across domains. By leveraging the additive information from modulation spectrograms and SSL embeddings, the fusion model demonstrates robustness to dataset variations and offers a promising direction toward generalizable fake speech detection.

\subsection{Generalization across language}
We analyze the behavior of both the baseline and proposed fusion models across individual languages in the MLAAD evaluation set, under two training conditions: using only the ASVspoof 2019 dataset (monolingual English) and using the combined ASVspoof 2019 and MLAAD datasets (multilingual). The evaluation protocol includes bonafide speech from four languages—German, Spanish, Russian, and Ukrainian—and fake speech across $23$ languages. For each fake language, we compute the EER using its scores as false and all bonafide scores as true. Fig.~\ref{fig:radar_plot} presents these language-wise EERs in a radar chart, excluding seven languages with fewer than $100$ fake samples.

Plot (a) shows the results for models trained on ASVspoof 2019, where the proposed fusion model consistently outperforms the baseline across all evaluated languages, suggesting better generalization in out-of-domain scenarios. Plot (b) presents the performance when models are trained on the combined ASVspoof 2019 and MLAAD datasets. The overall EERs are lower than in plot (a), indicating the benefits of multilingual training. The fusion model continues to show improved results for most languages, with performance comparable to the baseline in Maltese, Finnish, and Romanian. These findings suggest that the proposed fusion model provides improved language robustness for fake speech detection, showing better generalization in both cross-lingual and multilingual training scenarios compared to the SSL-AASIST baseline.

\section{Conclusion} \label{sec:conclusion}

This paper presents a novel approach for improving domain generalization in FSD by fusing modulation spectrogram feature with SSL embeddings. The proposed fusion leverages additive information providing a more generalizable representation. Integrated with the AASIST back-end, the (SSL+MS)-AASIST model outperforms the SSL-AASIST baseline in both in-domain and most out-of-domain evaluations. Additionally, the model demonstrates enhanced language robustness in multilingual scenarios. Future work will focus on exploring the integration of additional features and advanced training strategies for further performance improvement.

\section*{Acknowledgment}

This work was supported by MeitY, RCI Hyderabad, and SERB, India through various projects. Jagabandhu Mishra was supported by Academy of Finland (``SPEECHFAKES").

\newpage

\bibliographystyle{IEEEtran}
\bibliography{mybib.bib}


\end{document}